\documentclass[journal]{IEEEtran}

\ifCLASSINFOpdf
\else
\fi

\usepackage{graphicx}
\usepackage{epstopdf}
\usepackage{multicol}
\usepackage{url}
\usepackage{amsmath}
\usepackage{amssymb}
\usepackage{esint}
\usepackage{amsfonts}
\usepackage{multirow}
\usepackage{wrapfig}
\usepackage{cite}
\usepackage{subcaption}
\usepackage{lipsum}
\usepackage{float}	
\usepackage{algorithm}
\usepackage{algorithmic}
\usepackage[none]{hyphenat}
\usepackage[usenames,dvipsnames]{color}

\begin{document}

\title{Occupancy Counting with Burst and Intermittent Signals in Smart Buildings\thanks{This work was made possible by the National Science Foundation Grant AST\textendash1443999. The statements made herein are solely the responsibility of the authors.} \thanks{Bekir S. Ciftler and Kemal Akkaya are with the Department of Electrical and Computer Engineering, Florida International University, Miami, FL, USA (e-mail: $\lbrace$bcift001; kakkaya$\rbrace$@fiu.edu).} \thanks{ Sener Dikmese and Ismail Guvenc are with the Department of Electrical and Computer Engineering, North Carolina State University, Raleigh, NC, USA (e-mail: $\lbrace$sdikmes; iguvenc$\rbrace$@ncsu.edu).}\thanks{Abdullah Kadri is with Qatar Mobility Innovation Center, Qatar University, Doha, Qatar (e-mail: abdullahk@qmic.com).}}
\author{\IEEEauthorblockN{Bekir Sait \c{C}iftler, \IEEEmembership{Student Member, IEEE}, Sener Dikmese, \IEEEmembership{Member, IEEE}, \.{I}smail G\"{u}ven\c{c}, \IEEEmembership{Senior Member, IEEE}, Kemal Akkaya, \IEEEmembership{Senior Member, IEEE}, and Abdullah Kadri, \IEEEmembership{Member, IEEE}
}\vspace{-0.8cm}}
% \author{\IEEEauthorblockN{Bekir Sait \c{C}iftler\IEEEauthorrefmark{1}, Sener Dikmese\IEEEauthorrefmark{7}, Abdullah Kadri\IEEEauthorrefmark{4}, Kemal Akkaya\IEEEauthorrefmark{1}, and
% \.{I}smail G\"{u}ven\c{c}\IEEEauthorrefmark{7}}
% \IEEEauthorblockA{\\\IEEEauthorrefmark{1}Dept. of Electrical and Computer Engineering, Florida International University, Miami, FL, USA}
% \IEEEauthorblockA{\\\IEEEauthorrefmark{7}Dept. of Electrical and Computer Engineering, North Carolina State University, Raleigh, NC, USA}
% \IEEEauthorblockA{\\\IEEEauthorrefmark{4}Qatar Mobility Innovations Center (QMIC), Qatar University, Doha, Qatar}}
\maketitle
\begin{abstract}
	
		Zone-level occupancy counting is a critical technology for smart buildings and can be used for several applications such as building energy management, surveillance, and public safety.
		Existing occupancy counting techniques typically require installation of large number of occupancy monitoring sensors inside a building as well as an established network.	
		In this study, in order to achieve occupancy counting, we consider the use of WiFi \emph{probe requests} that are continuously transmitted from WiFi enabled smart devices.
		To this end, WiFi Pineapple equipment are used for \emph{passively} capturing ambient probe requests from WiFi devices such as smart phones and tablets, where no connectivity to a WiFi network is required. This information is then used to localize users within coarsely defined occupancy zones, and subsequently to obtain occupancy count within each zone at different time scales. An interacting multi-model Kalman filter technique is developed to improve occupancy counting accuracy. 
		Our numerical results using WiFi data collected at a university building show that utilization of WiFi probe requests can be considered a viable solution tool for zone-level occupancy counting in smart buildings.
\end{abstract}


\section{Introduction}
\label{sect:Introduction}

Smart cities of the future are expected to provide better use of public resources, increase quality of service offered to citizens, and reduce operational costs of public administrators~\cite{6917563}. Internet of Things (IoT) technology is a key enabler for smart cities, and it can support a plethora of services, ranging from building health inspection, to waste management, noise/air quality monitoring, traffic congestion control, city energy consumption reduction, smart parking/lighting, and automation of smart buildings~\cite{6740844,6702523,6525602}. Realizing the vision of smart cities necessitate effective use of IoT technologies for proximity detection, localization, tracking of objects and humans, and occupancy monitoring~\cite{7355289,6525604}.

\begin{figure}[t]
	\centering
	\includegraphics[width=1\columnwidth]{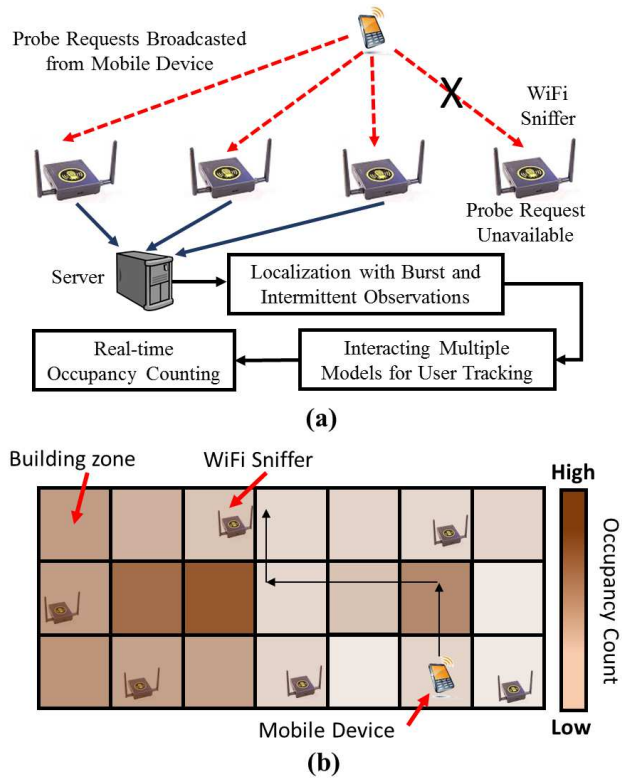}
	\caption{Occupancy counting using probe requests: (a) Capturing burst and intermittent probe requests at multiple WiFi Pineapples; (b) Zone-based real-time occupancy counting.}
	\label{fig:ExpModel}
    \vspace{-5mm}
    \end{figure}
    
Smart buildings constitute a key component of smart cities, which will benefit extensively from the use of IoT technologies for health monitoring, energy management, public safety, and surveillance,~\cite{6827323}. In particular, buildings are among the largest consumers of electricity in the United States: they account for $40\%$ of primary energy consumption and $72\%$ of electricity consumption\cite{6917563}.
An important portion of the electricity consumption of buildings is used for heating, ventilation, and air conditioning (HVAC). To this end, low cost, seamless, and accurate occupancy counting can help in achieving significant energy savings in smart buildings, such as by dynamically scheduling HVAC activity based on real-time building occupancy levels at different areas~\cite{agarwal2010occupancy}.

Occupancy counting in smart buildings can be implemented via video processing and camera systems or deployment of occupancy sensors throughout the building~\cite{akkaya2015}. 
These options require installation of new equipment that are often costly to deploy.
An alternative way is to use ambient wireless signals of opportunity that uniquely match to building occupants. 
While other technologies such as sensors, cameras, and RFID provide occupancy tracking at certain building zones, WiFi technology is already available in most buildings, and can provide good occupancy monitoring coverage.
Even though several positioning solutions exist based on available WiFi infrastructure, they typically require a connection between user equipment and WiFi access points~(APs)~\cite{akkaya2015}\cite{Balaji2013sentinel}.

In this paper, we consider the use passive sniffing of WiFi \emph{probe requests} for occupancy monitoring and tracking in smart buildings as illustrated in Fig.~\ref{fig:ExpModel}.
Probe requests are signals that are continuously broadcast from devices with WiFi technology, such as smartphones, laptops, and tablets\cite{edwin2016indoor,demir2013wi,Freudiger:2015:TYM:2766498.2766517,Musa:2012:TUS:2426656.2426685,namiot2014analysis}.
The probe requests are not encrypted, and can be captured and decoded with the help of passive sniffers as shown in Fig.~\ref{fig:ExpModel}(a), without connecting to a particular network. 
Probe requests are also burst in nature, since they are broadcasted in the air in search of WiFi networks to get connected, to get a list of available networks, or to handover between WiFi APs. 
Frequent transmission of probe requests from mobile devices introduces an opportunity to track the occupancy count in different zones of a building by simply monitoring the probe requests (see Fig.~\ref{fig:ExpModel}(b)). 
%Probe requests' burst and intermittent nature creates challenges for tracking since it needs to be smoothed for removing noise and outliers.
In particular, we can capture the received signal strength (RSS) of probe requests using sniffers such as WiFi Pineapple (WiFi-PA)~\cite{hak5}, which can then be used for occupancy monitoring inside the building. 

%On the other hand, burst and intermittent nature of probe requests introduce challenges for occupancy tracking, and effective pre-processing techniques are needed, to 1) remove noise and outliers from large number of other signals which are not building occupants, and 2) extract the periodic RSS information of each user from burst measurements.

To the best knowledge of the authors, there are no detailed studies in the literature that report efficiency of occupancy counting using WiFi probe requests using varying number of reference nodes to track individual devices. 
In this paper, we use WiFi probe requests captured at various reference locations for occupancy monitoring in smart buildings as summarized in Fig.~\ref{fig:ExpModel}. 
To this end, seven WiFi-PAs are deployed at various locations within the FIU Engineering Center, %(see Fig.~\ref{fig:FIUECMAP}), 
and probe requests are collected over multiple days.
The burst and intermittent nature of probe requests require post processing of the data to make it ready for the localization and tracking. 
Subsequently, at every sampling window, various localization techniques are used to obtain location estimates of each WiFi device, which are further refined using an Interacting Multiple Model (IMM) filter for tracking user location. 
Position estimates are then aggregated into one of the eight occupancy zones inside the building for real-time occupancy counting. 
%Our numerical results show that it is possible to effectively track the occupancy count using probe requests in different zones inside a building which can be utilized in various smart building applications.

The main novel contributions of this paper can be summarized as follows
	\begin{itemize}
    	\item We propose an adaptive tracking algorithm, which enables to study occupancy count with burst and intermittent measurements, as well as the varying number of positioning reference nodes (sniffers) during location tracking. To authors' best knowledge, there is no similar study considering both conditions in the literature.
		\item We develop estimators and heuristic methods to localize and track the target node even with measurements from a single reference node in the worst case. 
        In our experiments, we observe that $47.3\%$ of the time, probe requests are detected by a single reference node whereas $36.9\%$ of them are received by two reference nodes.
        Thus, including them in tracking and occupancy counting is crucial for robustly estimating occupancy count in building zones.
        \item We implement simulations with realistic WiFi channel models to show the occupancy detection performance of our proposed framework. We also show the accuracy of our approach using real world data, by carrying out an experiment with WiFi-PAs as the probe request sniffers for occupancy tracking for an indoor university campus environment. 
        Proposed method achieves up to $90\%$ performance in zone-level tracking.
        Our simulation results agree with experimental results, which show close performance.
	\end{itemize} 
% \begin{figure}[t]
% 	\centering
% 	\includegraphics[width=1\columnwidth]{Figures/OccupMesh.jpg}
% 	\caption{Occupancy counting is done considering partitioned mesh grid based on probe request RSS of devices.}
% 	\label{fig:OccupMesh}
%     \vspace{-5mm}
%     \end{figure}

This paper is organized as follows. In Section~\ref{sect:LiteratureReview}, a brief overview of the existing literature related to the tracking techniques is presented.
Models for the tracking algorithm are explained in Section~\ref{sect:SystemModel}.
Localization and initialization for user tracking using probe request data are presented in Section~\ref{sect:localization}, while tracking with the IMM filters and zone-level occupancy counting are provided in Section~\ref{sect:IMM}. Subsequently, 
gathering of probe requests is explained in Section~\ref{sect:probe} in detail.
Simulation and experimental results are presented in Section~\ref{sect:NumericalResults}, and finally, concluding remarks and future prospects are summarized in~Section~\ref{sect:Conclusion}.

\section{Literature Review}
\label{sect:LiteratureReview}

Use of probe requests have recently received interest from researchers for various applications. For example, they are used for load balancing in wireless networks to find hidden and mobile nodes in~\cite{ryou2010probe}, and to analyze the handover processes of 802.11 network in~\cite{mishra2003empirical}.
A WiFi flood attack detection system, which is a method based on the probe request and probe response timeouts, is proposed in~\cite{millikenl2012}. Privacy issues with probe requests are evaluated in \cite{cunche2012iknow}, where authors utilize probe requests to link the devices by creating a table for the requested WiFi network names and comparing them with the other devices.

Although RSS-based tracking with Kalman filters (KF) is a well-established area, it is still an interesting research topic due to its various new applications for accurate localization~\cite{zampella2015indoor,dardari2015indoor}.
In~\cite{castro2014simultaneous} two-slope RSS model is used with two Extended Kalman Filters (EKF) considering an IMM framework to improve tracking accuracy. In~\cite{yao2015exponential}, the authors present a novel exponential-Rayleigh RSS model for device-free localization and tracking with KF, where the main contribution is to include multipath components in RSS model for improving localization and tracking accuracy.
The accuracy is shown to increase significantly compared to the standard RSS model.
In~\cite{yungeun2012smartphone}, the RSS variance problem in tracking due to the hardware differences, device placement, and environmental changes are studied, and a particle filter based solution is proposed.
The results show that the tracking accuracy is more robust against the RSS variance when the accelerometer and digital compass readings are included in system.
In \cite{au2013indoor}, a novel approach based on compressive sensing is used for the localization and tracking of WiFi devices. In particular, coarse estimates of RSS fingerprinting are improved by the compressive sensing techniques, and KF with map information is used to track the devices.

In the papers mentioned above, a constant number of reference nodes are always assumed to continuously localize and track the users. However, in practice, reference nodes may fail to detect transmitted signal or may not be available at all times.
Even though there are several studies suggesting increased localization and tracking accuracy by introducing secondary reference nodes based on their location estimates in cooperative networks~\cite{gholami2012improved}, or self-adaptive localization techniques with a similar approach~\cite{dil2012rss}, they have not investigated localization with the limited and dynamically changing number of the reference nodes.

% Another challenge in the use of probe requests is their burst and intermittent transmission characteristics.
% In~\cite{shi2016Extended}, the authors consider a stochastic Bernoulli switching sequence to model intermittent signals. 
% They propose an algorithm to mitigate with the problem of intermittency by adapting measurement model to varying conditions.

% \begin{figure}[t]
% 	\centering \includegraphics[width=0.65\columnwidth]{Figures/Localization_Technologies}
% 	\caption{Occupancy monitoring using various different IoT technologies. Darker colors represent a better occupancy monitoring accuracy, and white colors represent no occupancy monitoring in particular building zone.}
% 	\label{fig:Occup_Monitoring}
%     \vspace{-5mm}
% \end{figure}
% \begin{figure}[H]
% 	\centering
% 	\caption{Occupancy tracking using probe requests that are captured at multiple WiFi Pineapples deployed in a building.}
% 	\label{fig:ToyModel3}
%     \vspace{-5mm}
%     \end{figure}

%     \begin{figure}[H]
% 	\centering
% 	\caption{Occupancy tracking using probe requests that are captured at multiple WiFi Pineapples deployed in a building.}
% 	\label{fig:ToyModel2}
%     \vspace{-5mm}
%     \end{figure}

\section{System Model}
\label{sect:SystemModel}

\begin{figure}[t]
	\centering
	\includegraphics[width=1\columnwidth]{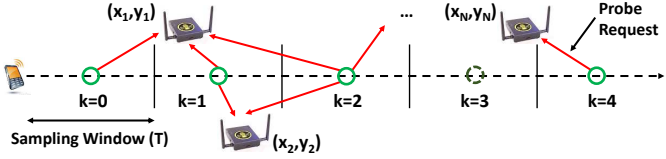}
	\caption{Tracking by using varying number of reference nodes.}
	\label{fig:ToyModel}
    \vspace{-5mm}
    \end{figure}
    
\begin{figure}[t]
	\centering
	\includegraphics[width=1\linewidth]{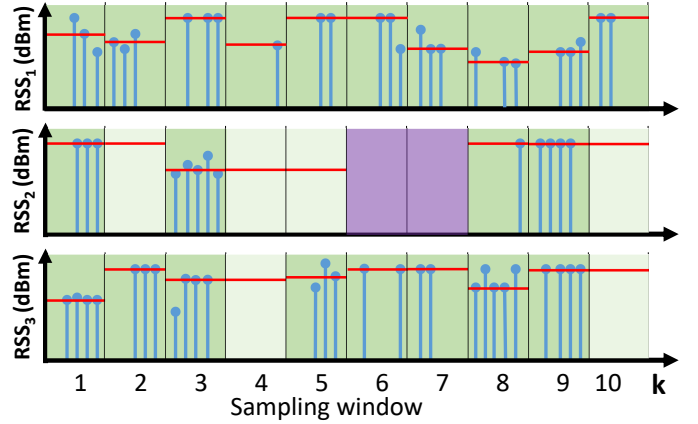}
	\caption{Burst and intermittent probe requests from a mobile device, observed at three different WiFi-PAs within different sampling windows, with sample-and-hold window duration $L=2$.}
	\label{fig:SignalModel}
    \vspace{-5mm}
\end{figure}

In this paper, we consider an occupancy counting scenario as seen in Fig.~\ref{fig:ToyModel}. There are $N$ reference nodes $\mathcal{R}~=~\{r_1,\cdots,r_N\}$ placed through the tracking zone. 
The known location of $i$th reference node is denoted with $\boldsymbol{x}_i=(x_i, y_i)$, and a target node has an unknown position $\boldsymbol{x'}=(x',y')$.
Burst and intermittent probe requests are broadcasted from the mobile target node as shown in Fig.~\ref{fig:SignalModel}.
We consider a predefined sampling window, where data from all the burst probe requests within the window are aggregated. 
In particular, the $k$th sampling window of duration $T$ spans the interval between time instances between $t_k$ and $t_{k-1}$.
Eventually, the true location of the target node in the sampling window $k$ is represented with $\boldsymbol{x'}_k=(x'_k,y'_k)$.

During tracking of the mobile target node, some reference nodes may not detect transmitted probe requests by the target node due to being far away from the target, poor link quality, and collision of probe request packets, as illustrated in Fig.~\ref{fig:ToyModel}. 
In such cases, we need to be able to work with RSS from the rest of the reference nodes, which are called {\textit{available measurements}} in sampling window $k$, and denoted by a set $\mathcal{P}_k=\{p_{i,k},\ldots,p_{u,k}\}$, where $p_{i,k}$ represents RSS value at reference node $i$ at sampling window $k$.
% The representative
% RSS values for the probe requests received during the sampling window $k$ at all the available reference nodes $\mathcal{R}_k~=~\{r_i,\ldots,r_j\}$ are captured by  set $\mathcal{P}_k = \{p_{i,k},\ldots,p_{j,k}\}$,
% where $p_{i,k}$ represents RSS value at reference node $i$ at sampling window $k$. 

We refer $p_{i,k}$ as a representative RSS value for window $k$ at node $i$, since there might be multiple probe requests observed from a certain target node within the same sampling window as seen in Fig.~\ref{fig:SignalModel}. Then, we have to extract the representative RSS value from these burst of observations.
Even though the RSS values are usually close or equal to each other for spatio-temporally close probe requests, there might be outliers due to the varying channel conditions.
Using the mean of the RSS values would therefore bias our results towards outliers; therefore, median value of the RSS values within the sampling window is used as a representative RSS for that sampling window as follows
\begin{align}
p_{i,k}=\mathrm{median}~\{p_{i,k,1},\ldots,p_{i,k,m_{i,k}}\}\label{eq:medianrec},
\end{align}
where the $i$th reference node is considered to receive $m_{i,k}$ different probe requests within the sampling window $k$. Let the RSS measurement for the probe request with the median RSS value be defined as 
\begin{align}
p_{i,k}=P_0 - 10n\log_{10}\bigg(\dfrac{d_{i,k}}{d_0}\bigg)+w, 
\label{eq:RSS}
\end{align}
where $P_0$ is the signal strength at the reference distance $d_0$, $n$ is the path loss exponent (PLE), $d_{i,k}=\|\boldsymbol{x}'_{k}-\boldsymbol{x}_{i}\|$ is the Euclidian distance between the $i$th reference node and the target device at sampling window $k$,
and $w\sim\mathcal{N}(0,\sigma^2)$ is the noise modeled by additive white Gaussian noise (AWGN) with a variance of $\sigma^2$ and mean of zero.

Once the representative RSS measurements are obtained for each reference node $i$ within the $k$th sampling window, the number of the available representative RSS measurements from the different reference nodes at sampling window $k$ is given by $|\mathcal{P}_k |$, which is the cardinality of set $\mathcal{P}_k$.
Due to the intermittent nature of probe requests that may result in unavailable measurements within subsequent measurement windows, a sample-and-hold approach is used to hold RSS values for a certain number of sampling windows at different reference nodes.
Otherwise, the useful ranging information from the difference nodes would be lost.
In Fig.~\ref{fig:SignalModel}, an example sampling windows is given with $L=2$, where $L$ is the holding length.

Assume that the last probe request detected by the $i$th reference node is within the sampling window $k$, and the $i$th reference node did not receive any probe requests within future sampling window(s). 
Then, the RSS value of the $i$th reference node for the samples after the sampling window $k$ using a holding length $L$ can be written as
\begin{align}
\mbox{$
	p_{i,k+l}=
	\begin{cases}
    p_{i,k+l},~&  \text{if } m_{i,k+l}\geq1\\
    p_{i,k},~&  \text{if } \sum_{s=1}^{l} m_{i,k+s}=0
	\end{cases}~$,
}\label{eq:RSShold}
\end{align}
% \begin{align}
% p_{i,k+l}=p_{i,k} \mbox{ if } m_{i,k+l}=0~\mbox{for}~l=1,\ldots,L.
% \label{eq:RSShold}
% \end{align}
for $l=1,\cdots,L$, given $m_{i,k}\geq1$.
If there is still no available signal received at the $i$th reference node after the sample-and-hold window, the RSS measurement will be removed from the set of the available RSS values $\mathcal{P}_k$ which represented with purple color in Fig.~\ref{fig:SignalModel}.

Given this system model and framework, our main goal in this paper is to determine zone occupancy based on tracking position of WiFi devices with varying number of the available reference nodes and intermittent transmission of the probe requests. In the following sections, we explain the localization of the individual target nodes, IMM-based tracking with varying conditions, and finally zone-level mapping and occupancy counting of WiFi devices.

\begin{figure}[t]
	\centering
	\includegraphics[width=1\linewidth]{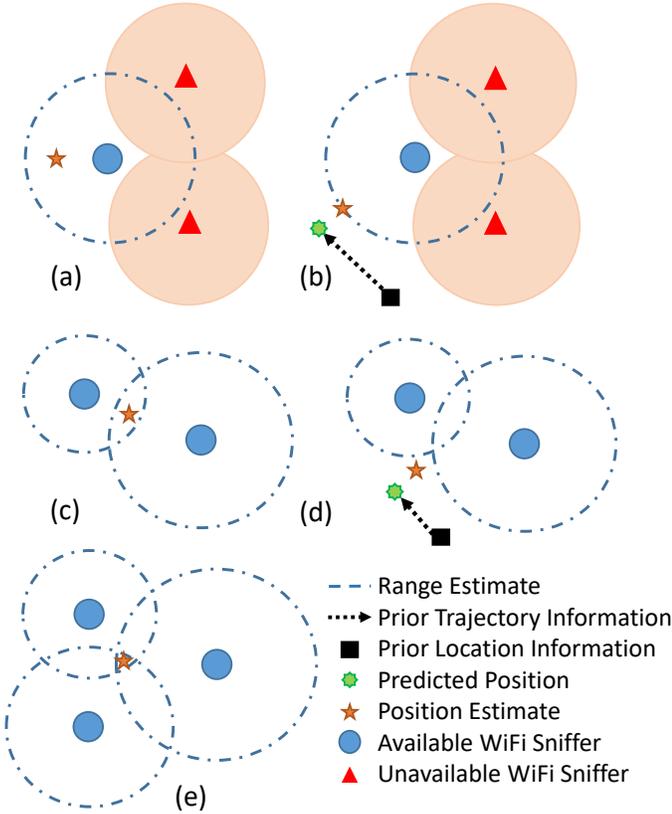}
	\caption{Tracking scenarios under the different circumstances, (a)~initialization with $|\mathcal{P}_k|=1$ and (b) Model-1, (c)~initialization with $|\mathcal{P}_k|=2$ and (d) Model-2, and (e)~both Initialization with $|\mathcal{P}_k|\geq3$ and Model-3.}
	\label{fig:SystemModel2}
    \vspace{-5mm}
\end{figure}

\section{Localization and Initialization Techniques}
\label{sect:localization}

In this section, RSS-based localization of the target node with varying number of reference nodes and different levels of a~priori information will be presented.
The technique used for localization changes with the number of available reference nodes and the unknown parameters is shown in Table~\ref{Table:Models}.
Since the unknown parameters are revealed in time, we will use them in the following sampling windows as prior information.
The novelty of our system lies in the use of any number of reference nodes without discarding any single probe request. 

Below, we will first consider heuristics and algorithms for estimating target location when we do not know the prior location information of the target (e.g., the target enters a building and no prior measurements available). 
Subsequently, in Section~\ref{sect:IMM} we will investigate how we can improve the localization accuracy when the prior trajectory of the target is available.

\subsection{RSS Available at Three or Less Reference Nodes}\label{Sec:ThreeOrLess}

First, consider the case that the RSS measurement in the sampling window $k$ is available at only a single reference node (i.e. when $|\mathcal{P}_k|=1$), and $P_0$ and $n$ are unknown.
Since $P_0$ and $n$ can only be estimated when $|\mathcal{P}_k|\geq4$ (see Section~\ref{Sec:FourOrMore}), we consider that (as in~\cite{Bose2007Practical}) $P_0$ and $n$ are obtained by averaging over (potentially limited number of) ground truth measurements where the location of the target is known.
\begin{table}[t]
\centering
\caption{Tracking models under various circumstances.}
\label{Table:Models}
\begin{tabular}{ |c|c|c| }
\hline
& No Bayesian Tracking & Bayesian Tracking\\
 \hline
  $|\mathcal{P}_{k}|=1$& Heuristic-1 & Model-1 ($j=1$) \\
 \hline
$|\mathcal{P}_{k}|=2$& Heuristic-2& Model-2 ($j=2$)\\
 \hline
 $|\mathcal{P}_{k}|=3$& Heuristic-3 & Model-3 ($j=3$)\\
 \hline
 $|\mathcal{P}_{k}|\geq4$& NLS & Model-3 ($j=3$)\\
 \hline
\end{tabular}
\end{table}
Then, the target device is assumed to be at the vicinity of the available reference node, and we may estimate the location of the target device to be 
\begin{align}
\boldsymbol{\hat{x'}}_k=\mathrm{median}~\{\mathcal{S}_u-\cup_{u=1}^N \mathcal{S}_u|u\neq i\},\label{eq:locest1}
\end{align}
where $\mathcal{S}_i$ is the set of possible locations in the coverage of $i$th reference node, considering a circular coverage area for all reference nodes.
In other words, \eqref{eq:locest1} implies that the location is estimated to be in the median location of possible positions sampled densely within the coverage area of the reference node.
On the other hand, as shown in Fig.~\ref{fig:SystemModel2}(a), coverage area of several reference nodes may overlap, and the location of the target node is assumed not to be within the coverage area of the reference nodes that do not have the RSS measurement from the target node.
Even though the location estimate in~\eqref{eq:locest1} is relatively coarse and based solely on the coverage area of the reference nodes, it is acceptable for the zone-level estimation, and we can still utilize the probe request measurements even if they are available only at a single reference node. It also allows to initialize the IMM tracking algorithm in Section~\ref{sect:IMM} at a better location for quicker and more accurate estimation, and to avoid local optimums.

Second, consider that the RSS information from the target node at sampling window $k$ is available only at two reference nodes ($|\mathcal{P}_k|=2$) and no prior information is available on the location of the target node. 
Then, the location estimate $\boldsymbol{\hat{x}}_k$ for the sampling window $k$ can be estimated as follows
\begin{align}
\boldsymbol{\hat{x'}}_k=\boldsymbol{x}_i+\bigg(\hat{d}_{i,k}+\frac{\hat{d}_{i,k}+\hat{d}_{u,k}-d_{iu}}{2}\bigg)\dfrac{\boldsymbol{x}_u-\boldsymbol{x}_i}{\|\boldsymbol{x}_u-\boldsymbol{x}_i\|},\label{eq:locest2}
\end{align}
where $d_{ij}$ represents the distance between the $i$th and $u$th reference nodes, and
the estimated distance between the $i$th reference node and the target node ($\hat{d}_{i,k}$) based on the RSS measurements is given as
\begin{align}
\hat{d}_{i,k}=10^{\frac{P_0-p_{i,k}}{10n}}.\label{eq:Rang}
\end{align}
In other words, as shown in~Fig.~\ref{fig:SystemModel2}(c), the target node is assumed to be on the straight line that passes through the locations of the two reference nodes, and positioned on the line depending on the RSS measurements at each node.

Third, consider that at sampling windows $k$, RSS measurements are available at three or more reference nodes ($|\mathcal{P}_k|\geq3$).
Then, we consider a linear least squares (LLS) solution as a low computational complexity method~\cite{guvenc2012fundamental,guvenc2009survey} that has acceptable localization performance as shown in Fig.~\ref{fig:SystemModel2}(e).
The location of the target device can be obtained by solving the LLS problem in the form $\boldsymbol{{\rm M}x}_k=\boldsymbol{\rm b}$ with
\begin{align}
\boldsymbol{\rm M}=&
  \begin{bmatrix}
    x_1-\dfrac{1}{N}\sum^{N}_{i=1}x_i & y_1-\dfrac{1}{N}\sum^{N}_{i=1}y_i\\
    \vdots & \vdots \\
    x_N-\dfrac{1}{N}\sum^{N}_{i=1}x_i & y_N-\dfrac{1}{N}\sum^{N}_{i=1}y_i
  \end{bmatrix},\\
  {\rm b}=\dfrac{1}{2}&\begin{bmatrix}
(x_1^2-\dfrac{1}{N}\sum^{N}_{i=1}x_i^2)+(y_1^2-\dfrac{1}{N}\sum^{N}_{i=1}y_i^2)\\-(\hat{d}_{1,k}^2-\dfrac{1}{N}\sum^{N}_{i=1}\hat{d}_{i,k}^2)\\
\vdots\\
(x_N^2-\dfrac{1}{N}\sum^{N}_{i=1}x_i^2)+(y_N^2-\dfrac{1}{N}\sum^{N}_{i=1}y_i^2)\\-(\hat{d}_{N,k}^2-\dfrac{1}{N}\sum^{N}_{i=1}\hat{d}_{i,k}^2)
  \end{bmatrix},
\end{align}
where $\boldsymbol{\rm M}$ and $\boldsymbol{\rm b}$ are defined with the known position of the reference nodes ($x_i,y_i$) and ranging information for reference nodes using \eqref{eq:Rang}.
The solution to the LLS problem can then be found by solving for $\boldsymbol{x'}_k$, as follows
\begin{align}
\boldsymbol{\hat{x'}}_k=(\boldsymbol{\rm M}^T\boldsymbol{\rm M})^{-1}\boldsymbol{\rm M}^T\boldsymbol{\rm b}.\label{eq:locest3}
\end{align}

\subsection{RSS Available at Four or More Reference Nodes}\label{Sec:FourOrMore}

Consider now that the RSS information is available at least at four reference nodes, i.e., $|\mathcal{P}_k|\geq 4$. Then, adapting \cite{gholami2013rss}, a nonlinear least squares algorithm is used to solve the localization problem with unknown parameters.
The optimization problem is based on the relaxation of a maximum likelihood estimator with Taylor expansions of RSS defined in \eqref{eq:RSS}, around assumed values of unknown parameters\cite{gholami2013rss}.
The general optimization problem is defined as a special type of quadratic problem, general trust region problem, given as follows
\begin{align}
\mathop {\rm minimize}\limits_{ {\boldsymbol{y}}_{v}} ~~ &\Vert  {\boldsymbol{\rm Z}}_{v}  {\boldsymbol{y}}_{v}-  {\boldsymbol{b}}_{v}\Vert^{2}\label{eq:generallocalization}\\
{\rm subject~to} \quad& { \boldsymbol{y}}_{v}^{T} {\boldsymbol{D}}_{v} {\boldsymbol{y}}_{v}+2 { \boldsymbol{f}}^{T}_{v} { \boldsymbol{y}}_{v} =0,\nonumber
\end{align}
where $\boldsymbol{\rm Z}_v$ is the translation matrix and $\boldsymbol{y}_v$ is the vector for unknown parameters and their derivations for case $v$, and $\boldsymbol{b}_v$ holds the measurements regarding the unknown parameters.
The constraint ${ \boldsymbol{y}}_{i}^{T} {\boldsymbol{D}}_{i} {\boldsymbol{y}}_{i}+2 { \boldsymbol{f}}^{T}_{i} { \boldsymbol{y}}_{i} =0$ defines the feasible set for solution of optimization, where ${\boldsymbol{D}}_{i}$ and ${\boldsymbol{f}_i}$ are the selection diagonal  and vector, respectively.
Each of these vectors and matrices are defined and given explicitly for three different availability level of a~priori information below (i.e., estimated or unknown $P_0$ and $n$).

\paragraph{Unknown $P_0$ ($v=1$)} In the case of unknown $P_0$ and $\boldsymbol{x}'_k$ with knowledge of $n$, at least three reference nodes are required to estimate $P_0$.
If the number of the reference nodes are less than three, approaches introduced in Section~\ref{Sec:ThreeOrLess} are used.
The optimization problem given in \eqref{eq:generallocalization} for this case is written as
\begin{align}
&\boldsymbol{\rm Z}_1\triangleq\begin{bmatrix}
\lambda_1 &-2\lambda_1\boldsymbol{a}_1&-1\\
\vdots & \vdots & \vdots\\
\lambda_N &-2\lambda_N\boldsymbol{a}_N&-1 
\end{bmatrix},\boldsymbol{f}_1\triangleq\bigg[
-\frac{1}{2}~0~0~0\bigg]^T,\nonumber\\
&\boldsymbol{b}_1\triangleq\big[-\lambda_1\|\boldsymbol{a}_1\|^2,\cdots,-\lambda_N\big\|\boldsymbol{a}_N\|^2]^T,\boldsymbol{D}_1\triangleq \textnormal{diag}(0,1,1,0),\nonumber
\end{align}
where $\lambda_i\triangleq10^{p_{i}/(5n)}$ and $\alpha=10^{P_0/(5n)}$, and solution for minimization problem is defined as $\boldsymbol{y}_1\triangleq\big[\|\boldsymbol{x}'_k\|^2~\boldsymbol{x}'^T_k~\alpha\big]$.

\paragraph{Unknown $n$ ($v=2$)} In the case of unknown $n$ and $\boldsymbol{x}'_k$ with knowledge of $P_0$, optimization matrices and vectors become
\begin{align}
&\boldsymbol{\rm Z}_2\triangleq\begin{bmatrix}
1 &-2\boldsymbol{a}_1&q_1\ln q_1\\
\vdots & \vdots & \vdots\\
1 &-2\boldsymbol{a}_N&q_N\ln q_N
\end{bmatrix},\boldsymbol{f}_2\triangleq\bigg[
-\frac{1}{2}~0~0~0\bigg]^T,\nonumber\\
&\boldsymbol{b}_2\triangleq\big[q_1-\|\boldsymbol{a}_1\|^2,\cdots,q_N-\|\boldsymbol{a}_N\|^2\big]^T,\boldsymbol{D}_2\triangleq {\rm diag}(0,1,1,0),\nonumber
\end{align}
where $q_i\triangleq10^{(P_0-p_{i,k})/(5n_0)}$ and $n_0$ is the tuning parameter of the algorithm for $n$. 
The final solution of the optimization is defined as $\boldsymbol{y}_2\triangleq\big[\|\boldsymbol{x}'_k\|~\boldsymbol{x}'^T_k~\delta\big]$ where $\delta\triangleq\frac{(n-n_0)}{n_0}$ shows the suitability of the tuning parameter.

\paragraph{Unknown $P_0$ and $n$ ($v=3$)} In the worst case, which there is no prior information on the parameters, we have to estimate both the location of the target nodes and channel parameters with below matrices and vectors
\begin{align}
&\boldsymbol{\rm Z}_3\triangleq\begin{bmatrix}
1 &-2\boldsymbol{a}_1&-g_1&g_1\ln g_1\\
\vdots & \vdots & \vdots& \vdots\\
1 &-2\boldsymbol{a}_N&-g_N&g_N\ln g_N
\end{bmatrix},\nonumber\\
&\boldsymbol{b}_3\triangleq\big[-\|\boldsymbol{a}_1\|^2,\cdots,-\|\boldsymbol{a}_N\|^2\big]^T,\nonumber\\
&\boldsymbol{f}_3\triangleq\bigg[
-\frac{1}{2}~0~0~0~0\bigg]^T,~\boldsymbol{D}_3\triangleq {\rm diag}(0,1,1,0,0),\nonumber
\end{align}
where $g_i\triangleq10^{(\bar{P_0}-{p_{i,k}})/(5n_0)}$ and $\bar{P_0}$ is the tuning parameter for $P_0$. Solution for the corresponding NLS to a general trust region problem is defined as $y_3\triangleq\big[\|\boldsymbol{x}'_k\|^2~\boldsymbol{x}'^T_k~\gamma~\gamma\delta\big]^T$ where $\gamma\triangleq10^{(P_0-\bar{P_0})/(5n)}$ is a measure on $P_0$ tuning parameter suitability.
The IMM based tracking takes the localization estimates as an input, which is explained in the following section in details.

\section{Interacting Multiple Model Tracking and~Occupancy Counting}
\label{sect:IMM}

In this section, we study Bayesian tracking of the individual WiFi devices by exploiting prior location information of the mobile device. 
Probe requests are intermittent, which means that the target device may not send any probe signals over a duration of many sampling windows. Even when the probe requests are transmitted by the mobile device, they may not always be captured at the reference node. Therefore, using a standard Kalman filtering technique (see e.g. \cite{guvenc2003enhancements,yim2008extended}) that assumes uniform set of system parameters will not work effectively. 
Considering varying system properties such as the number of the available measurements and prior information related to the device, we consider an IMM framework that employs multiple, interacting Kalman filters in this study.
The overall model~\cite{hartikainen2011optimal} is represented with a linear system as follows 
\begin{align}
&\boldsymbol{\rm x}_k=\boldsymbol{\rm A}^j_{k-1}\boldsymbol{\rm x}_{k-1}+\boldsymbol{\rm q}^j_{k-1}\\
&\boldsymbol{\rm y}_k=\boldsymbol{\rm H}^j_k\boldsymbol{\rm x}_k+\boldsymbol{\rm r}_k^j,
\end{align}
where $\boldsymbol{\rm x}_k\in\mathbb{R}^n$ is the state of the system during sampling~window~$k$, $\boldsymbol{\rm A}_{k-1}^j$ is the transition matrix for the model order $j$, where the model order depends on the number of available reference nodes (see Table~\ref{Table:Models}). which is in effect during the sampling windows $k-1$, and $\boldsymbol{{\rm q}}^j_{k-1}\sim{\mathcal{N}(\boldsymbol{0},\boldsymbol{\rm Q}^j_{k-1}})$ is the process noise which is Gaussian distributed with zero mean and covariance $\boldsymbol{\rm Q}^j_{k-1}$ for the $j$th model. 
The measurement on the sampling window $k$ is denoted by $\boldsymbol{\rm y}_k~\in~\mathbb{R}^m$, where $\boldsymbol{\rm H}_k^j$ is the $j$th measurement model matrix, and $\boldsymbol{\rm r}_k^j~\sim~\mathcal{N}(\boldsymbol{0},\boldsymbol{\rm R}^j_{k})$ is the measurement noise on the step $k$ for $j$th model with Gaussian distribution of mean zero and covariance $\boldsymbol{\rm R}_{k}^j$.
As explained earlier, each model in the system has its own Kalman filter parameters, and the resulting estimate is calculated with the selection of the correct model based on the status of the system such as $|\mathcal{P}_k|$ and prior information of channel parameters.

A standard Wiener process velocity model is used for all of the models, with varying measurement noise $(\boldsymbol{\rm r}_k^j)$ and step sizes based on sampling window length $(\Delta t_k=T)$.
Defining matrices of the models are given as
\begin{align}
\boldsymbol{\rm A}^j_k=&\begin{bmatrix}
1 & 0 & \Delta t_k & 0\\
0 & 1 & 0 & \Delta t_k\\
0 & 0 & 1 & 0\\
0 & 0 & 0 & 1
\end{bmatrix},~
\boldsymbol{\rm H}^j_k=&\begin{bmatrix}
1 & 0 & 0 & 0\\
0 & 1 & 0 & 0
\end{bmatrix}.
\end{align}

\subsection{Review of a Generic Kalman Filter}\label{Sec:GenericKF}

Kalman filtering, in general, has two steps: 1) prediction and 2) update. In the prediction step, next state of the system is predicted with the given previous measurements, and in the update step, the current state of the system is estimated with the given measurement at that sampling window~\cite{hartikainen2011optimal}.
In below, first the general procedure for the prediction and update stages are mathematically summarized (model index $j$ will be dropped for brevity), and we will subsequently discuss how Kalman filters with multiple models can be utilized for our scenario.  

\paragraph{Prediction}
\begin{align}
\boldsymbol{\rm m}_k^-&=\boldsymbol{\rm A}_{k-1}\boldsymbol{\rm m}_{k-1}~,\label{eq:predicted}\\
\boldsymbol{\rm U}_k^-&=\boldsymbol{\rm A}_{k-1}\boldsymbol{\rm U}_{k-1}\boldsymbol{\rm A}_{k-1}^T+\boldsymbol{\rm Q}_{k-1}~,
\end{align}{
where $\boldsymbol{\rm m}_k^-$ and $\boldsymbol{\rm U}_k^-$ are the predicted mean vector and the predicted covariance matrix of the state in sampling window~$k$, respectively, while $\boldsymbol{\rm m}_{k-1}$ and $\boldsymbol{\rm U}_{k-1}$ are the estimated mean vector and the estimated covariance matrix of the state in sampling window~$k-1$.
Mean and covariance of states consist of position and velocity information of the target device.}

\paragraph{Update}
\begin{align}
\boldsymbol{\rm v}_k&=\boldsymbol{\rm y}_k-\boldsymbol{\rm H}_k\boldsymbol{\rm m}_k^-~,\\
\boldsymbol{\rm S}_k&=\boldsymbol{\rm H}_k\boldsymbol{\rm U}^-_k\boldsymbol{\rm H}_k^T+\boldsymbol{\rm R}_k~,\\
\boldsymbol{\rm K}_k&=\boldsymbol{\rm U}_k^-\boldsymbol{\rm H}_k^T\boldsymbol{\rm S}_k^{-1}~,\\
\boldsymbol{\rm m}_k&=\boldsymbol{\rm m}_k^-+\boldsymbol{\rm K}_k\boldsymbol{\rm v}_k~,\\
\boldsymbol{\rm U}_k&=\boldsymbol{\rm U}_k^--\boldsymbol{\rm K}_k\boldsymbol{\rm S}_k\boldsymbol{\rm K}_k^T,
\end{align}
where $\boldsymbol{\rm v}_k$ and $\boldsymbol{\rm S}_k$~are the measurement residual and measurement prediction covariance on state for sampling window~$k$, respectively. 
Kalman gain $\boldsymbol{\rm K}_k$ is the tuning parameter, which adjusts prediction's correction for sampling~window~$k$.
Estimated mean and covariance of the state at sampling window~$k$ with adding the measurements are represented with $\boldsymbol{{\rm m}_k}$ and $\boldsymbol{\rm U}_k$, respectively. 

\subsection{IMM Based Kalman Tracking of Target Location}
\label{Sec:IMMBasedKT}

Section~\ref{Sec:GenericKF} summarized the model for a generic Kalman filter; in this section, we consider a multi-modal Kalman filter, whose parameters change with respect to the model order, and the models are selected based on the number of available RSS measurements at different reference nodes. 
These models also vary by their measurement error since the number of measurements affects the localization accuracy. Each filter receives a raw location estimate as an input (relying solely on measurements), and outputs a filtered location output using a technique that varies by the model.

First, consider that the RSS of the probe request from the target mobile device is available only at a single reference node (i.e. $|\mathcal{P}_k|=1$). Then, since we have access to prior location estimate, a better location estimate than that was introduced in \eqref{eq:locest1} and Fig.~\ref{fig:SystemModel2}(a) can be obtained, as summarized in Fig.~\ref{fig:SystemModel2}(b) as Model-1 ($j=1$). {The target node is assumed to be on the straight line between the predicted position for sampling window $k$ (represented by $\boldsymbol{x}_k^-$),
and the position of the available reference node ($\boldsymbol{x}_i$), utilizing ranging information from the reference node as
\begin{align}
\boldsymbol{\hat{x}}'_k=\boldsymbol{x}_i+\hat{d}_{i,k}\dfrac{\boldsymbol{x}'^-_k-\boldsymbol{x}_i}{\|\boldsymbol{x}'^-_k-\boldsymbol{x}_i\|}~, 
\end{align}
where $\hat{d}_{i,k}$ is as in~\eqref{eq:Rang}.
In other words, position of target node is estimated as $\hat{d}_{i,k}$ far from available reference node (from measurements), towards predicted position by IMM.}
In particular, the predicted position based on previous location estimate and trajectory are used for reducing the error of the estimation in Fig.~\ref{fig:SystemModel2}(a) which relied only on a single reference node measurement.

When $|\mathcal{P}_k|=2$ (probe RSS measurement available at two reference nodes), a similar approach as illustrated in Fig.~\ref{fig:SystemModel2}(d) called as Model-2 is followed.
In particular, we consider the predicted position of the target node ($\boldsymbol{x}'^-_k$) in \eqref{eq:predicted} for compensating bias with estimated location in \eqref{eq:locest2} as 
\begin{align}
\boldsymbol{\hat{x}}'_k=\dfrac{\boldsymbol{x}'^-_k+\boldsymbol{x}_i+\bigg(\hat{d}_{i,k}+\frac{\hat{d}_{i,k}+\hat{d}_{u,k}-d_{iu}}{2}\bigg)\dfrac{\boldsymbol{x}_u-\boldsymbol{x}_i}{\|\boldsymbol{x}_u-\boldsymbol{x}_i\|}}{2}.
\end{align}
{Simply, the position of the target node is first estimated as in \eqref{eq:locest2} using the measurements from the two reference nodes, and then the bias of the estimation is adjusted with the predicted position by taking average of the estimate and the prediction.}

Finally, in Model-3 (i.e. when $|\mathcal{P}_k|\geq3$), LLS or NLS estimator is used based on the number of reference nodes and unknown parameters.
If either $P_0$ or $n$ is unavailable, we use NLS to estimate the location as well as that parameter, as described in Section~\ref{Sec:FourOrMore}.
If there are no unknown parameters (i.e. all the parameters are estimated), LLS is used to estimate the location due to its lower computational complexity.

\begin{figure}[t]
	\centering
	\includegraphics[width=1\linewidth]{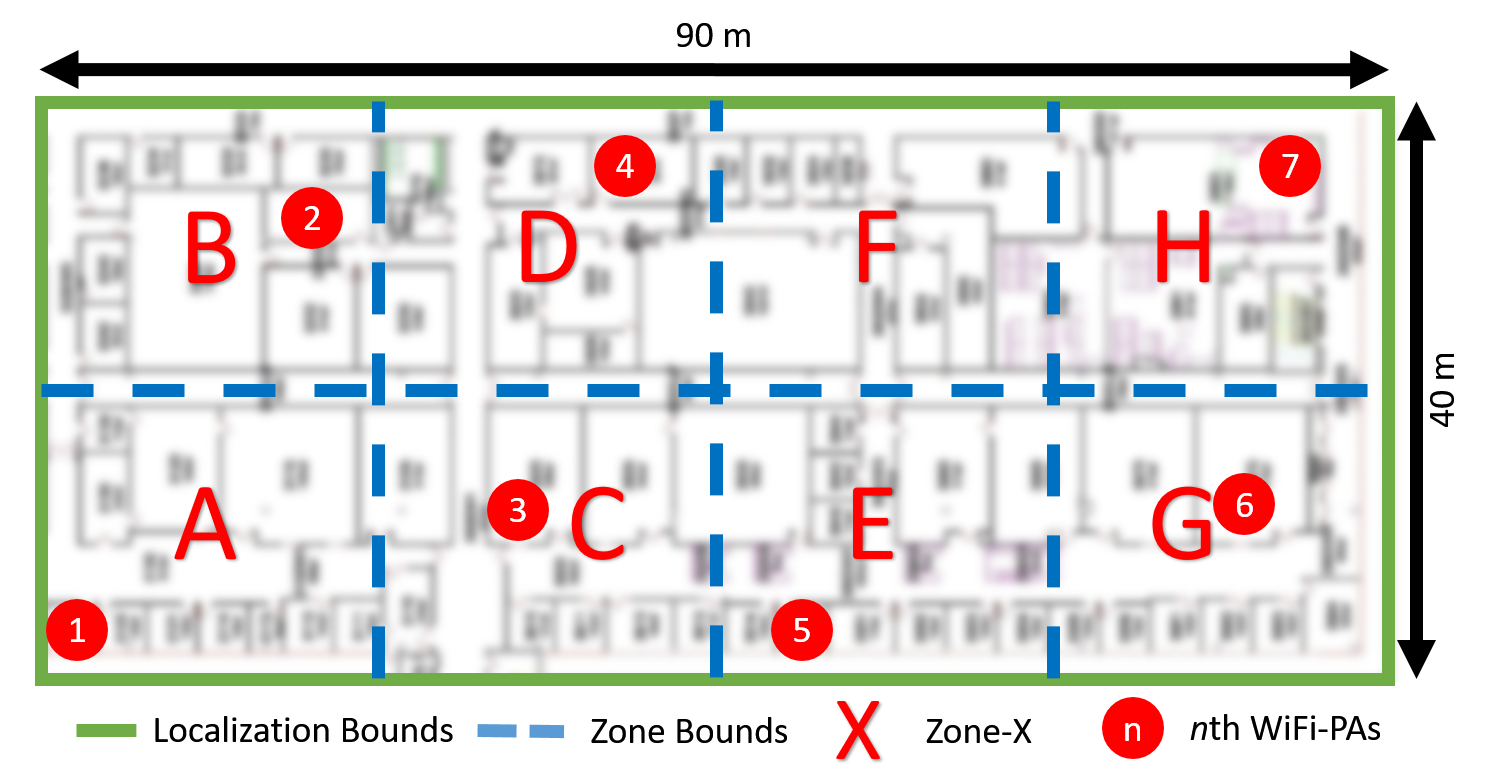}
	\caption{Map of building floor used in experiments with eight different building zones, as well as the locations of the WiFi sniffers placed across the different zones.}
	\label{fig:FIUECMAP}
        \vspace{-5mm}
\end{figure}

{

\subsection{Interaction of Multiple Models}
IMM consists of three major steps: interaction
(mixing), filtering and combination. 
In each sampling window $k$, the initial
conditions are obtained for certain model, from the measurement and previous sampling window with the knowledge of $|\mathcal{P}_k|$, hence right model is selected as given in Table~\ref{Table:Models}.
Then standard Kalman filtering is performed for each model, and after a weighted combination of updated
state estimates produced by all the filters yielding a final estimate for the state ($\boldsymbol{\rm m}_k$) and covariance of the state ($\boldsymbol{\rm U}_k$) in that particular sampling window $k$. 
The weights are chosen according to the probabilities of the models, which is binary and deterministic in our case since $|\mathcal{P}_k|$ is known. Thus, $\boldsymbol{\rm m}_k$ and $\boldsymbol{\rm U}_k$ are updated for IMM filter as follows

\begin{align}
\mbox{$
	\boldsymbol{\rm m}_k=
	\begin{cases}
    \boldsymbol{\rm m}_k^1,~&  \text{if } j=1\\
    \boldsymbol{\rm m}_k^2,~&  \text{if } j=2\\
	\boldsymbol{\rm m}_k^3,~&  \text{if } j=3
	\end{cases}$,~$
	\boldsymbol{\rm U}_k=
	\begin{cases}
    \boldsymbol{\rm U}_k^1,~&  \text{if } j=1\\
    \boldsymbol{\rm U}_k^2,~&  \text{if } j=2\\
	\boldsymbol{\rm U}_k^3,~&  \text{if } j=3
	\end{cases}~.$
}\label{eq:L}
\end{align}

}

In each model, the $2\times 2$ measurement noise covariance matrix $\boldsymbol{\rm R}_k^j$ varies with the number of measurements as follows
\begin{align}
\mbox{$
	\boldsymbol{\rm R}^j_k=
	\begin{cases}
    \mbox{diag}(\sigma_M^2/1,\sigma_M^2/1),~&  \text{if } j=1\\
    \mbox{diag}(\sigma_M^2/2,\sigma_M^2/2),~&  \text{if } j= 2\\
	\mbox{diag}(\sigma_M^2/4,\sigma_M^2/4),~& \text{if } j=3
	\end{cases}~,$
}\label{eq:L}
\end{align}
where $\sigma_M^2$ represents the variance of the error in the location estimates. Using the covariance matrix as in~\eqref{eq:L} for different models ensures that the prediction is weighted more when there are limited number of available reference nodes, while measurements are weighted more when there are at least three or more available reference nodes.

\begin{figure}[t]
	\centering
	\includegraphics[width=1\linewidth]{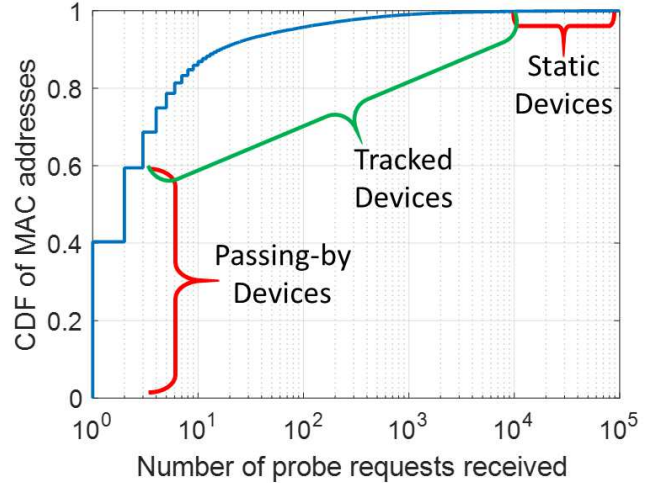}
	\caption{Distribution of passing-by, static and tracked devices by number of probe requests.}
	\label{fig:SamplePAs}
        \vspace{-5mm}
\end{figure}

% \textcolor{blue}{Since number of available reference nodes is known at every sampling window $k$, models are selected deterministically.}
\subsection{Occupancy Counting}
\label{Sect:OccupancyCounting}
Our final goal is real-time occupancy monitoring, by counting the number of users within coarsely defined occupancy zones in the area of interest. 
The IMM filtering results lead to mapping of the each target node to a zone which is created as illustrated in Fig.~\ref{fig:ExpModel}.
Sum of the individual target nodes in a zone gives the occupancy count of that zone.
In this case, there might be a problem when a user switches the device's screen off since probe requests may not be sent for $300$ seconds to $600$ seconds\cite{internshipreport2013}.
Thus, this interval should be considered as a grace period for that individual device, and the user should be kept within the occupancy count for that zone, until the time threshold expires.
In our experimental model, we consider the grace period to be $300$ seconds, after which we remove a user from the building zone which has been last localized.

\section{Sniffing Probe Requests}
\label{sect:probe}    

	In the experiments, seven different WiFi-PA equipment are deployed at various locations on the floor of a university campus building as in Fig.~\ref{fig:FIUECMAP}.
% 	WiFi-PA has dual integrated radios that comes with the Atheros AR9331 system on a chip (SoC), which includes a 400~MHz MIPS processor, 16~MB ROM and 64~MB RAM.
% 	Busybox architecture which is based on Linux is used to capture WiFi probe request packets.
The data captured at WiFi-PA include absolute time stamp providing the time at which the data was captured, MAC address of the WiFi device, and the RSS of the WiFi device. Gathered data is transferred to a server for execution of the occupancy counting.

Probe request mechanism is explained in \cite{mishra2003empirical}.
Probe requests, which are sent by WiFi devices to scan WiFi APs at certain channels, are active mechanisms to discover APs around the mobile device.
APs respond with probe responses and beacons.
A problem with probe requests is passing-by and static devices.
For instance, there might be people outside of the building passing by a WiFi device and it still can be detected by the reference nodes which are close to the border.
Such devices usually send a few low power probe requests while in the vicinity of the reference node.
Static devices (such as wireless printers) send thousands of probe requests with the same RSS value usually periodically.
Passing-by and static devices are removed from the occupancy count in our analysis using outlier detection and filtering techniques. 
Our experimental results indicate that only $40\%$ of the detected devices are in our region of interest as shown in Fig.~\ref{fig:SamplePAs}.
In total, $33,847$ unique and legitimate MAC addresses are detected within a period of one week.

Since probe requests are only used to discover APs, they are burst and intermittent as shown in Fig.~\ref{fig:SignalModel}.
After connecting to a network, probe requests are rarely sent to search for a network with better quality or handover possibilities.
Therefore, they could be sent 50 times within a minute (i.e., burst), or do not trigger for 5 to 10 minutes (i.e., intermittent).
This depends on the implementation of the IEEE 802.11 protocol at the device in terms of both the design of hardware and software.
Due to the mobility, WiFi device needs to send probe requests more frequently since it causes to lose connectivity with the AP, or degrade the quality of the connection.
Therefore, probe requests can effectively facilitate localization and tracking of the WiFi devices.

\begin{table}[t]
	\centering
	\caption{Simulation parameters.}
	\begin{tabular}{cc}
		\hline
		\multicolumn{1}{|l|}{\textbf{Parameter}} & \multicolumn{1}{l|}{\textbf{Value}} \\ \hline\hline
		\multicolumn{1}{|l|}{WiFi protocol} & \multicolumn{1}{l|}{IEEE 802.11n} \\ \hline
		\multicolumn{1}{|l|}{Floor width and length} & \multicolumn{1}{l|}{$40$ m x $90$ m} \\ \hline
		\multicolumn{1}{|l|}{{Sampling window length (T)}} & \multicolumn{1}{l|}{{$3$ seconds}}  \\ \hline
        \multicolumn{1}{|l|}{{Sampling window hold length (L)}} & \multicolumn{1}{l|}{{$1$ to $15$}}  \\ \hline
        \multicolumn{1}{|l|}{{Occupancy grace period}} & \multicolumn{1}{l|}{{$5$ minutes}}  \\ \hline
		\multicolumn{1}{|l|}{{Number of the reference nodes}} & \multicolumn{1}{l|}{{$7$}}  \\ \hline
        \multicolumn{1}{|l|}{{Path loss exponent ($n$)}} & \multicolumn{1}{l|}{{$2-6$}}  \\ \hline
        \multicolumn{1}{|l|}{{Reference distance ($d_0$)}} & \multicolumn{1}{l|}{{$1$ meter}}  \\ \hline
        \multicolumn{1}{|l|}{{Signal strength at reference distance ($P_0$)}} & \multicolumn{1}{l|}{{$-35$ dBm}}  \\ \hline
        \multicolumn{1}{|l|}{{Noise floor}} & \multicolumn{1}{l|}{{$-90$ dBm}}  \\ \hline
        \multicolumn{1}{|l|}{{Detection threshold}} & \multicolumn{1}{l|}{{$-90$ dBm}}  \\ \hline
		\label{table:parameter}
            \vspace{-5mm}
	\end{tabular}
\end{table}

\section{Numerical Results}
\label{sect:NumericalResults}

\begin{figure}[t]
	\centering
	\includegraphics[width=1\linewidth]{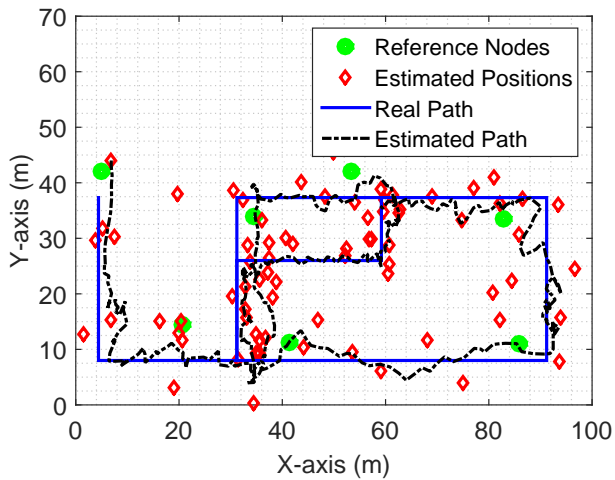}
	\caption{A sample simulation for a given map (Fig.~\ref{fig:FIUECMAP}).}
	\label{fig:TrackingVisual}
        \vspace{-5mm}
\end{figure}

In this section, we will first investigate the performance of the proposed framework with computer simulations using realistic WiFi signal propagation data. Subsequently, we will present our experimental results and occupancy counting based on the measurement data that has been collected at FIU College of Engineering and Computing building. 

\subsection{Simulation Results}

The proposed framework is simulated considering an indoor environment with propagation parameters and indoor channel characteristics summarized in Table~\ref{table:parameter}.
In our simulations, we used a realistic multipath indoor channel model to generate the RSS values. 
It is compared with the ideal flat fading channel results, where there is only a single strong path together with the AWGN.
Probe request statistics {such as interarrival times and detection probability} are included from experimental measurements as well to increase the reliability of the results.
By introducing detection probability, detection of probe requests by single or multiple reference nodes is modeled realistic.
Also, path loss exponents and reference received powers are used from the obtained measurements.

\begin{figure}[t]
	\centering
	\includegraphics[width=1\linewidth]{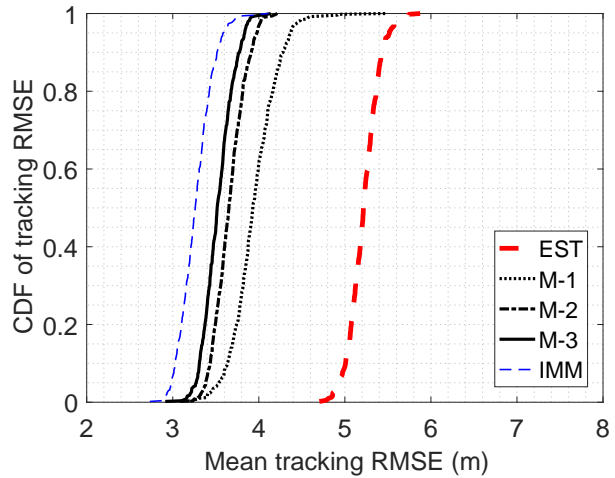}
	\caption{CDF of mean tracking RMSE for the simulation.}
	\label{fig:MeanTrackingRMSE}
        \vspace{-5mm}
\end{figure}

\begin{figure}[t]
	\centering
	\includegraphics[width=1\linewidth]{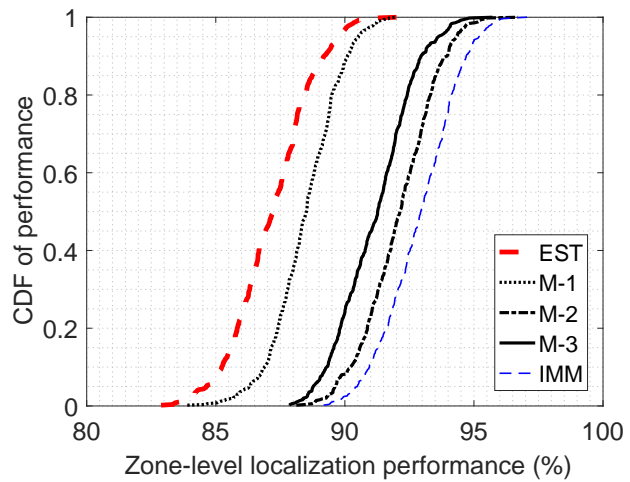}
	\caption{Performance of zone-level tracking for simulation.}
	\label{fig:ZoneLevel}
        \vspace{-5mm}
\end{figure}

\begin{figure}[t]
 	\centering
	\includegraphics[width=1\linewidth]{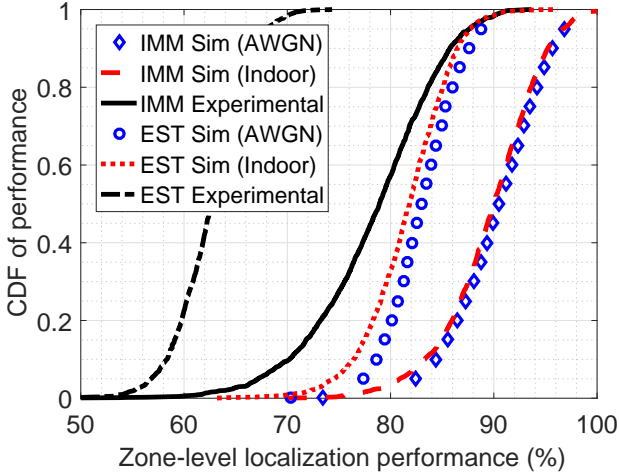}
	\caption{Simulation and experiment performance comparison of zone-level localization with and without~IMM.}
	\label{fig:ComparisonESTIMM}
    \vspace{-5mm}
\end{figure}

\begin{figure}[t]
 	\centering
	\includegraphics[width=1\linewidth]{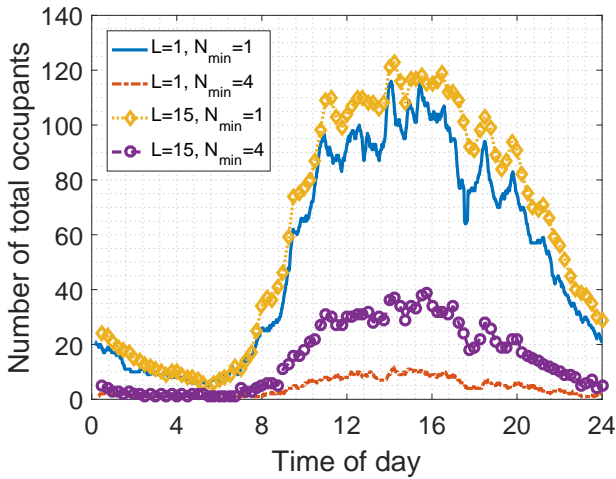}
	\caption{Total number of occupants on a weekday for sample-and-hold window lengths $L=1$ and $L=15$. Scenario with $N_{\rm min}=4$ necessitates at least four reference nodes to localize a target device, while $N_{\rm min}=1$ can localize a target device within a building zone using a single reference node.}
	\label{fig:OccupancyCountTotal}
        \vspace{-5mm}
\end{figure}

An example trajectory from the simulations are plotted as seen in Fig.~\ref{fig:TrackingVisual} with the individual position estimates before filtering, as well as the IMM output as the estimated path.
The estimated path is very close to the real path as we expect. The mean RMSE for the estimated path is $3.83$ meters in this realization, while mean RMSE for individual localization estimates is $6.17$ meters which is significantly higher.
Cumulative distribution function (CDF) of the mean tracking RMSE for bare localization estimates without IMM filtering (EST), and the IMM filtering results are compared for a given scenario in Fig.~\ref{fig:MeanTrackingRMSE}.
Simulations are averaged over 1000 Monte Carlo realizations with the realistic indoor channel parameters.
Median RMSE for IMM filter is $3.59$ meters, whereas $4.01$, $3.72$, and $4.13$ meters for Model-1 (M-1), Model-2 (M-2), and Model-3 (M-3), respectively. It is also $5.97$ meters for the individual location estimates (EST), which is almost twice the RMSE obtained from the IMM filter.

This performance is also reflected to the zone-level localization performance (i.e., percentage of correctly localizing a target device within the building zone where its true location belongs) as shown in Fig.~\ref{fig:ZoneLevel}.
Even though the zone-level localization performance varies from $88\%$ to $97\%$ for IMM, the median performance is $92.5\%$. Individual Kalman filters are very close to each other with the median performance of $88.5\%$ with range of $80\%$ to $95\%$.
Similar to the tracking RMSE, performance of zone-level localization is much lower for individual location estimates as the median performance is $83.2\%$, with a range between $76\%$ to $87\%$.

\subsection{Experimental Results}

In addition to simulations, extensive experimental data are collected at FIU to validate the proposed methods with real measurements. First, several experimental walks are considered using WiFi devices to compare the simulation and experimental results of the ground-truth data on a fixed path within the building, using the reference nodes as shown in Fig.~\ref{fig:TrackingVisual}.
The track is walked at different times of the day along with several WiFi devices.
The comparison of the zone-level localization performance is presented in Fig.~\ref{fig:ComparisonESTIMM}.
The performance of simulations with the AWGN and realistic indoor channel are close to each other for both the individual location estimates and the IMM.
The median performance of the individual location estimates is $84\%$ under ideal AWGN, whereas it is $83\%$ under realistic indoor channel. 
Performance with ground-truth data is lower with $64\%$.
The IMM improves the median performance to $90\%$ in the simulation results, whereas it increases to $78\%$ in the experimental results with real data.
The growth in the experimental performance using IMM is substantial with $14\%$ increase.
Hence, the IMM plays a critical role in the zone-level localization of the individual target devices.

Once we localize all the target mobile users in building zones (filtered out from passing by users and other fixed devices as shown in Fig.~\ref{fig:SamplePAs}) using IMM, we can obtain the real-time occupancy count in different building zones using these location estimates. 
Using experimental data, occupancy counting results over a period of 24~hours are shown in Fig.~\ref{fig:OccupancyCountTotal} as a sum of total occupants over all the zones in a week day.
As expected, peak hours of the occupancy in the building is between $12$~PM and $6$~PM, where most of the classes are scheduled.
The lowest occupancy is around $5$~AM.
Fig.~\ref{fig:OccupancyCountTotal} also shows that if we enforce the localization and occupancy counting technique to use at least $N_{\rm min}=4$ reference nodes ($|\mathcal{P}_k|\geq4$), a significant portion of the occupancy is not detected. %In particular, while $15$ people are detected using the conventional methods at the peak hours, it is possible to detect around $120$ people in total using the proposed method. 
Therefore, the proposed techniques in Fig.~\ref{fig:SystemModel2} and the IMM method can help in significantly improving the occupancy counting.

\begin{figure}[t]
 	\centering
	\includegraphics[width=1\linewidth]{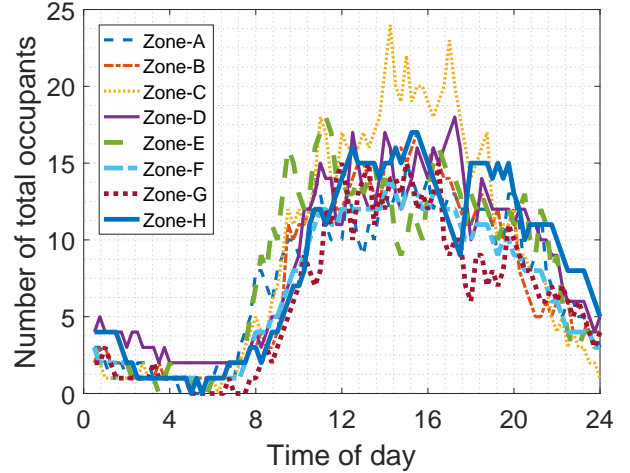}
	\caption{Total number of occupants within the individual zones given in Fig.~\ref{fig:FIUECMAP} throughout a weekday.}
	\label{fig:MultiZone24}
        \vspace{-5mm}
\end{figure}

\begin{figure}[t]
 	\centering
	\includegraphics[width=1\linewidth]{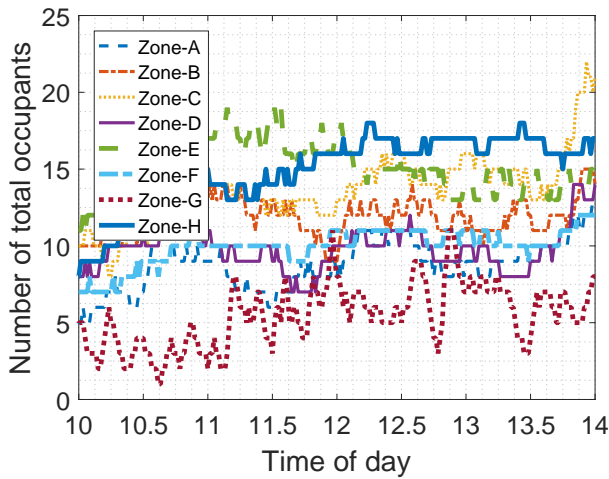}
	\caption{Total number of occupants within individual zones given in Fig.~\ref{fig:FIUECMAP} at peak hours.}
	\label{fig:MultiZone1014}
        \vspace{-5mm}
\end{figure}

Fig.~\ref{fig:OccupancyCountTotal} also compares the effects of the sample-and-hold window length $L$ on occupancy counting results. 
Even though the total occupancy count with $N_{\rm min}=4$ (i.e. the measurements are dropped if $|\mathcal{P}_k|\leq3|$) at peak hours increases to $40$ using a sample-and-hold method using holding length of $L=15$, this is still significantly lower compared to when we use IMM based tracking with $N_{\rm min}=1$ (i.e. measurements are used regardless of number of available reference nodes). Use of a larger holding window also smooths the total occupancy count when the IMM approach is used.

In Fig.~\ref{fig:MultiZone24} and Fig.~\ref{fig:MultiZone1014}, we provide occupancy counting results in each of the eight individual zones over a period of one whole week day and four hours, respectively, considering different time resolutions. 
Although the individual zones show similar behavior throughout the day, their peak hours are different as seen in Fig.~\ref{fig:MultiZone24}. For instance, Zone-E has a peak at 11~AM, while Zone-D has its peak at 5~PM.
The most common characteristic is that all of the zones have their minimum around 5~AM in the morning. Zone-C has the highest peak among all zones, with $24$ people around 3~PM.
Another interesting observation is that Zone-H, which consists a student study area and senior design laboratory, has occupants until very late hours, while Zone-C has least occupied zone at midnight, since it consists mostly administrative offices.
In Fig.~\ref{fig:MultiZone1014}, occupancy counts of individual zones are shown within peak hours, i.e. from 10~AM to 2~PM.
The movement of people from one zone to another is more visible in this figure.

\begin{figure}[t]
	\centering
	\includegraphics[width=1\linewidth]{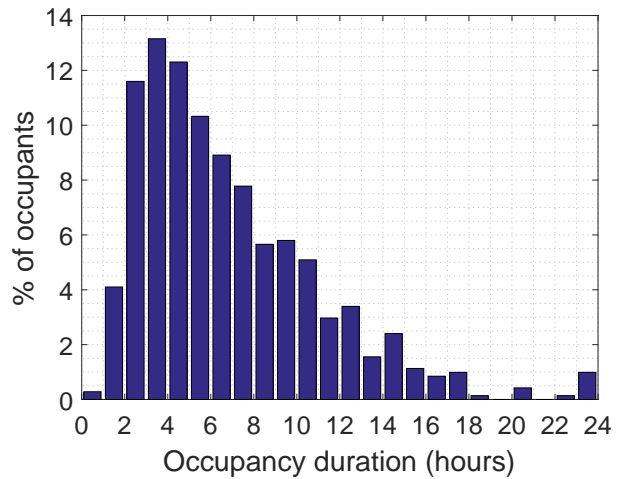}
	\caption{The distribution of the occupancy hours for the individual occupants.}
	\label{fig:OccupancyHours}
       \vspace{-5mm}
\end{figure}

In Fig.~\ref{fig:OccupancyHours}, percentage of occupants based on hours spent in the building is presented.
For instance, $11.5\%$ of the people spend $2$ to $3$ hours in the building, whereas the people spending less than $1$ hour is less than $1\%$.
More than half of the occupants spend less than $6$ hours in the building, with $13\%$ of them spending $3$ to $4$ hours.

\section{Conclusion}
\label{sect:Conclusion}

%In conclusion, probe requests are the good way to assess occupancy levels of a building on a zone-level basis.

In this paper, we propose a new framework for real-time occupancy counting of buildings by passively sniffing WiFi probe requests. We first differentiate the static and passing-by devices in the building using probe request statistics, and then use the information for the remaining users for localization, tracking, and occupancy counting inside the building. 
The proposed method assures the utilization of the probe requests regardless of the number of the available reference nodes that they are detected, which allows to detect three times more users in 
the occupancy counting process when compared to requiring at at least four reference nodes for localization.
In our experimental results, we are able to perform zone-level occupancy tracking with up to $90\%$ accuracy.
We are also able to determine the peak hours of the individual zones as well as the quiet times of the building. Our future work includes use of more advanced filtering techniques such as particle filters to further improve zone level localization and occupancy counting performance.

\section*{Acknowledgements}
We would like to thank Edwin Vattapparamban for helping in probe request data collection within Florida International University, and Yusuf Said Eroglu for reviewing the manuscript and providing feedback.
% This work was made possible by the National Science Foundation Grant AST-1443999. The statements made herein are solely the responsibility of the authors.

\bibliographystyle{IEEEtran}

\end{document}